\documentclass[aps,pra,pdf,superscriptaddress,twocolumn,showpacs,nofootinbib]{revtex4-2}
\usepackage{amsmath,amssymb,amsfonts}
\usepackage{eqnarray}
\usepackage{graphics,float}
\usepackage{bm}
\usepackage{braket}
\usepackage{amsmath}
\usepackage{physics}
\usepackage{mathtools}
\usepackage{gensymb}
\usepackage{color,xcolor,colortbl}
\usepackage{tikz}
\usepackage{soul}
\usetikzlibrary{shapes}
\usepackage[colorlinks=true,
linkcolor=blue,
filecolor=magenta,      
urlcolor=blue,
citecolor=blue]{hyperref}
\usepackage{lipsum}
\usepackage{orcidlink}
\usepackage[normalem]{ulem}

\definecolor{Gray}{gray}{0.85}
\definecolor{LightCyan}{rgb}{0.25,0.9,0.95}
\definecolor{LightMagenta}{rgb}{0.91,0.21,0.96}

\begin{document}

\title{Richtmyer-Meshkov instability induced by a Gaussian-kicked rotor-like potential in a three-component Bose-Einstein condensate}

\author{Susovan Giri  \orcidlink{0009-0003-3848-8662}}
\email{iamsusovangiri@gmail.com}
\affiliation{Department of Physics, Indian Institute of Technology Kharagpur, Kharagpur, West Bengal 721302, India}


\begin{abstract}
This research investigates the Richtmyer-Meshkov instability within a three-component Bose-Einstein condensate. We offer a kicking potential that replicates the impulsive shock across both interfaces of the three-component BEC to trigger the Richtmyer-Meshkov instability. We propose a hydrodynamic model that indicates the essential criteria for the onset of hydrodynamic instability in the BEC through the application of the hydrodynamic pressure equation. We utilise the split-step Crank-Nicolson method to numerically solve the coupled Gross-Pitaevskii equations, with the objective of obtaining the immiscible ground state of all components in a quasi-2D harmonic trap. Following this, we introduce a Gaussian-like disruption instigated by the kick rotor model at both interfaces of the confined three-component BEC to explore the time-dependent development of nonlinear hydrodynamic instability patterns, culminating in quantum turbulence and nonlinear configurations at the interfaces. The kinetic energy spectra of both the incompressible and compressible elements of the system exhibit scaling in the low-wavenumber domain with power-law exponents of $-5/3$ and $1$, respectively. Conversely, the high-wavenumber domain is defined by $k^{-4}$ and $k^{-7/2}$ scaling for the incompressible and compressible elements, respectively, aligning with the established spectral characteristics of quantum turbulence. This research offers fresh perspectives on the mechanisms of hydrodynamic instability, vortex generation, non-equilibrium pattern evolution, and quantum turbulence in superfluid systems comprising multiple interacting elements.

\end{abstract}

\maketitle

\section{Introduction}
\label{sec:intro}

The Richtmyer-Meshkov instability (RMI)~\cite{https://doi.org/10.1002/cpa.3160130207,meshkov_instability_1969,PhysRevE.102.013203,PhysRevFluids.4.103902,PhysRevA.82.043608} occurs when a shock wave interacts with the interface between two fluids of different densities. First identified by Richtmyer and later examined in greater detail by Meshkov, this instability is described as the impulsive analog~\cite{qpgn-6gg2} of the Rayleigh-Taylor instability (RTI)~\cite{https://doi.org/10.1112/plms/s1-14.1.170,10.1098/rspa.1950.0052,PhysRevA.80.063611,PhysRevA.85.013602,PhysRevA.81.053616,PhysRevA.108.013320,y226-17w9,PhysRevA.89.013631,PhysRevA.85.013630,doi:10.1126/sciadv.adw9752,PhysRevA.83.043623,PhysRevA.85.013602}. The key distinction between the two hydrodynamic instabilities lies in the temporal evolution of small perturbations: RTI leads to an exponential increase in perturbation amplitude, whereas RMI produces a linear growth in time. When the shock wave passes through the interface, it generates deformations on the perturbed surface, leading to the progressive growth of the initial perturbations. As the instability develops, the interface enters the nonlinear regime and evolves into characteristic structures such as spikes, bubbles, and mushroom-shaped patterns. 

Although the hydrodynamic instability has been extensively studied in classical fluids because of its relevance in inertial confinement fusion~\cite{betti_inertial-confinement_2016}, astrophysical processes~\cite{10.1063/1.874046}, and high-energy-density plasma research~\cite{PhysRevFluids.5.113701,paul_drake_approaches_2008} in different laboratory experiments~\cite{jacobs_experimental_1996,brouillette_experiments_1999,chapman_experiments_2006,dimonte_richtmyermeshkov_1996,PhysRevLett.70.1806,farley_high_1999,di_stefano_richtmyer-meshkov_2015,aglitskiy_perturbation_2006,Balakumar_Orlicz_Ristorcelli_Balasubramanian_Prestridge_Tomkins_2012,10.1063/1.3099320}, its realization in quantum fluids such as Bose-Einstein condensates (BECs) opens an exciting new frontier. In dilute-gas condensates, the immiscible interfaces between different hyperfine components or between distinct atomic species provide a controllable platform for studying shock-induced interfacial dynamics. In particular, a spatially localized quantum kick rotor model (QKR)~\cite{PhysRevLett.96.160403, PhysRevA.101.053625, PhysRevA.77.043621, PhysRevA.73.013601, PhysRevA.101.043624, PhysRevLett.100.024103, Mangaonkar_2020, K.Henderson_2006, PhysRevE.70.056206, PhysRevLett.92.054101} offers a versatile and experimentally feasible method for generating shock waves and driving nonequilibrium dynamics in multicomponent Bose-Einstein condensates. By imparting a transient momentum impulse to the condensate, the Gaussian kick~\cite{PhysRevA.82.043608, PhysRevA.83.043623} can trigger interfacial and hydrodynamic instabilities, including the Kelvin-Helmholtz instability (KHI)~\cite{PhysRevB.81.094517,PhysRevA.82.063604,PhysRevA.97.053608,PhysRevA.104.023312,1kgv-h6pw}, RTI and RMI, and induce complex vortex dynamics. The tunable amplitude and width of the kick provide a powerful tool to explore the interplay between shock-wave propagation, interface deformation, and the subsequent evolution of quantum turbulence (QT)~\cite{PhysRevA.106.023306, GHOSHDASTIDAR2022127776, pj6n-3w6p, 2bty-918t, sivakumar_energy_2024, PhysRevA.80.023618, PhysRevLett.119.184502} in phase-separated condensates. Quantum turbulence, as in classical fluids, displays an energy cascade from large to small scales, with the incompressible energy spectrum following $E(k)\propto k^{-5/3}$ in the inertial range. At large wave numbers a $k^{-4}$ Saffman scaling can emerge, which is related to random distribution of vortices in Bose-Einstein condensates (BEC)~\cite{GHOSHDASTIDAR2022127776, PhysRevA.80.023618, PhysRevLett.119.184502}. In contrast, the compressible spectrum $E_{\mathrm{c}}(k)$ is due to density fluctuations and phonon excitations, $E_{\mathrm{c}}(k)\propto k$ at long wavelengths and $E_{\mathrm{c}}(k)\propto k^{-7/2}$ at short wavelengths. 

Previous study~\cite{PhysRevA.82.043608} has demonstrated the occurrence of the RMI in the two-component BEC composed of two hyperfine states of $^{87}\mathrm{Rb}$. In their work, the instability is initiated by applying a magnetic pulse that drives the two condensate components toward each other. Depending on the strength of the magnetic pulse, several distinct dynamical regimes of the RM instability have been reported. Kobyakov et al.~\cite{PhysRevA.83.043623}  showed that the RTI is driven by a constant force, while the RMI is produced by a magnetic pulse, with each instability leading to characteristic growth of interfacial perturbations. Although the behaviour of the RMI is well documented in classical compressible fluids and plasmas, its manifestation in quantum fluids, particularly in three-component BECs, remains relatively unexplored. Despite extensive theoretical and numerical studies of RMI in binary and multicomponent BECs subjected to impulsive external perturbations, the role of a kick-rotor model in triggering this instability has not yet been investigated. Specifically, a comprehensive investigation of RMI in a three-component BEC remains lacking. Understanding how kick-induced driving and the resulting breaking of axial symmetry influence interfacial hydrodynamic instabilities is therefore an important open question. In this work, we address this question by investigating the development and nonlinear evolution of RMI in a three-component BEC subjected to a kick-induced perturbation.

Our study presents numerical simulations of the onset of hydrodynamic instabilities and the subsequent development of quantum turbulence in a phase-separated three-component BEC confined in a quasi-two-dimensional harmonic trap. Sufficiently large repulsive interatomic scattering lengths are employed to realize a fully phase-separated ground state, in which the second component is sandwiched between the first and third components, thereby forming two concentric circular interfaces. To trigger the instability, transient Gaussian kicks are applied independently to the first and second components, simultaneously perturbing both interfaces. A hydrodynamic framework is developed to analyze the resulting interface dynamics and characterize the evolution of the instability at the two boundaries of the condensate. We further investigate the emergence of quantum turbulence and demonstrate the appearance of the characteristic spectral scaling laws, namely $k^{-5/3}$, $k^{-4}$, $k$, and $k^{-7/2}$. 

This paper is organized as follows: Section~\ref{Theory} describes our theoretical foundation and the hydrodynamical model of our system. Section~\ref{Numerical} discusses the numerical results obtained for RMI. Section~\ref{QT} we analyze the phenomena of QT. Section~\ref{Conclusions} represents the conclusion of this study.

\section{Theory }
\label{Theory}
We want to describe the classical model of the instability alongside the Gross-Pitaevskii (GP) equation and the hydrodynamic model for quantum fluids such as BECs in order to identify and understand the key differences between the classical and quantum regimes. 

\subsection{Classical Model}
\label{Theory-A}
Richtmyer modelled classically RM instability based on Taylors work where he took gravity as Dirac delta function and proposed that the perturbation amplitude $\eta(t)$ is changed under shock wave as follows~\cite{https://doi.org/10.1002/cpa.3160130207,meshkov_instability_1969}: 
\begin{equation}
    \frac{\partial \eta(t)}{\partial t} = k A V \eta_0
\label{eqn::Richmayer}    
\end{equation}
where, $k$ is the wave number, V is the change in interface velocity, $\eta_0$ is the initial perturbation amplitude. The preshock atwood number A is defined as:
\begin{equation}
    A= \frac{\rho_1 -\rho_2}{\rho_1 + \rho_2} 
\label{eqn::Atwood}    
\end{equation}

where, $\rho_1$ and $\rho_2$ are the densities of the top and bottom layer, respectively. 

In the hydrdynamic instability, under a small sinusoidal perturbation of the interface, the perturbation amplitude $\eta(t)$ evolves as
\begin{equation}
\eta(t) = \eta_0 \, e^{\gamma t}
\end{equation}
where the linear growth rate $\gamma = \sqrt{A \alpha k}$ with $\alpha$ is acceleration.

From linear stability analysis, the dynamics of interfacial perturbations are governed
by the dispersion relation~\cite{osti_5063967}
\begin{equation}
\omega^2 = \frac{(\rho_1 - \rho_2) \alpha k + \sigma k^3}{\rho_1 + \rho_2},
\end{equation}
where $\sigma$ represents the interfacial tension coefficient. In the absence of perturbation, the frequency $\omega$ remains real for all values of $k$ and the dispersion relation describes capillary waves with frequency a characteristic scaling $\omega_c \propto k^{3/2}$, indicating stable capillary-wave dynamics.

\subsection{Kicked rotor BEC Model}
\label{Theory-B}

The Gross-Pitaevskii equation~\cite{PhysRevA.65.023603,PhysRevA.84.053623,PhysRevA.72.043613,PhysRevA.80.063621,pethick_boseeinstein_2008} for a trapped Bose-Einstein condensate with an additional
kicking potential is formulated by:

\begin{equation}
\begin{aligned}
i\hbar \frac{\partial \psi_j}{\partial t}
=
\Bigg[
-\frac{\hbar^2}{2m_j}\nabla^2
+ U_j
+ \sum_{j'=1}^{3} g_{jj'} |\psi_{j'}|^2
- V_j(r,t)\,
\Bigg]\psi_j .
\end{aligned}
\label{eqn::RNLS}
\end{equation}

Here $\psi_j$($j = 1, 2, 3$) is the macroscopic wave function with mass $m_j$ of the Rb atom. $U_j = \frac{1}{2} m_j \omega_r^2 \left( r^2 + \lambda^2 z^2 \right)$ is the circular symmetric trapping potential confirming the quasi-2D structure of the condensate considering $\lambda = \omega_z / \omega_r >> 1$, where \( \omega_r \) is the radial trapping frequency. The contact pseudopotential are characterized by the coupling strength $g_{jj'} = 2\pi \hbar^2 a_{jj'}/m_{jj'}$, where $a_{jj'}$ is the $s$-wave scattering length and $m_{jj'} = m_j m_{j'}/(m_j + m_{j'})$ is the reduced mass between atoms from components $j$ and $j'$. The kicking potential is described by $V_j(r,t)= \frac{r^2}{2} \phi_{d_j} cos(2k_{L_j} x) \delta(t)$, where $\phi_{d_j}$ is the kick strength and $k_{L_j}$ is the wave vector of the kick pulse for the $j$-th component.

\subsection{Hydrodynamic Model}
\label{Theory-C}

Here, we study the RMI in a three-component BEC composed of \(^{87}\mathrm{Rb}\) atoms, where a short-duration kick is used to perturb the interfaces between the components. We consider the three components to be completely immiscible, satisfying the condition \(g_{ij}^{2}>g_{ii}g_{jj}\). Initially, the system contains two circular equilibrium interfaces, \(S_1\) and \(S_2\), with radii \(r_1\) and \(r_2\), respectively. The interface \(S_1\) separates BEC-1 from BEC-2, while \(S_2\) separates BEC-2 from BEC-3. The corresponding interfacial tension coefficients are denoted by \(\sigma_1\) and \(\sigma_2\).

Following the variational formulation of interfacial dynamics in two-component BECs~\cite{PhysRevA.82.063604}, the Lagrangian for the present three-component system is written as~\cite{1kgv-h6pw}
\begin{equation}
\begin{aligned}
L={}&
\int d\theta
\Bigg[
\int_0^{r_1+\eta_1} rdrP_1
+\int_{r_1+\eta_1}^{r_2+\eta_2} rdrP_2
\\&\quad
+\int_{r_2+\eta_2}^{\infty} r\,dr\,P_3
\Bigg]
-\sigma_1S_1-\sigma_2S_2 ,
\end{aligned}
\label{eqn:Lagrangian}
\end{equation}
where
\begin{equation}
\begin{aligned}
P_j={}&
i\hbar\psi_j^*\frac{\partial\psi_j}{\partial t}
+\frac{\hbar^2}{2m_j}\psi_j^*\nabla^2\psi_j
\\
&-(U_j-V_j)|\psi_j|^2
-\frac{g_{jj}}{2}|\psi_j|^4 .
\end{aligned}
\label{eqn::Pressure}
\end{equation}
Here, \(U_j\) denotes the trapping potential and \(V_j\) represents the external kick potential. The two interfaces are parameterized as
\begin{equation}
R_1(\theta,t)=r_1+\eta_1(\theta,t),
\qquad
R_2(\theta,t)=r_2+\eta_2(\theta,t).
\end{equation}
Thus, BEC-1 occupies the region \(r<R_1\), BEC-2 occupies \(R_1<r<R_2\), and BEC-3 occupies \(r>R_2\).

The equations governing the interface dynamics are obtained by taking the functional derivatives of the action \(\int dt\,L\) with respect to the interface positions. This gives the pressure-balance conditions
\begin{equation}
P_1(R_1)-P_2(R_1)
+\frac{\sigma_1}{r_1^2}
\frac{\partial^2\eta_1}{\partial\theta^2}
=0,
\label{eqn::Pressure1}
\end{equation}
and
\begin{equation}
P_2(R_2)-P_3(R_2)
+\frac{\sigma_2}{r_2^2}
\frac{\partial^2\eta_2}{\partial\theta^2}
=0.
\label{eqn::Pressure2}
\end{equation}
These equations represent the linearized normal-pressure balance at the two interfaces, with the terms proportional to the interfacial tensions providing the restoring contribution.

We consider a stationary state in which the \(j\)-th component wavefunction is written as
\begin{equation}
\Psi_j=\sqrt{n_j(r)}\,e^{i\phi_j},
\label{eqn::Stationary wavefunction}
\end{equation}
where \(n_j(r)\) is the equilibrium density and \(\phi_j\) is the phase of the \(j\)-th component. The external potential \(U_j(r)\) and the equilibrium density are assumed to depend only on the radial coordinate. Within the sharp-interface approximation, the densities satisfy
\begin{equation}
n_1=0\quad\text{for}\quad r>R_1,
\qquad
n_2=0\quad\text{for}\quad r<R_1,
\end{equation}
and
\begin{equation}
n_2=0\quad\text{for}\quad r>R_2,
\qquad
n_3=0\quad\text{for}\quad r<R_2.
\end{equation}

For a homogeneous equilibrium system, the pressure balance at the two interfaces reduces to
\begin{equation}
\frac{g_{11}}{2}n_{s1}^{\,2}
=
\frac{g_{22}}{2}n_{s2}^{\,2},
\qquad
\frac{g_{22}}{2}n_{s3}^{\,2}
=
\frac{g_{33}}{2}n_{s4}^{\,2},
\label{eqn::equilibrium_pressure}
\end{equation}
where
\begin{equation}
n_{s1}=n_1(r_1^-),\qquad
n_{s2}=n_2(r_1^+),
\end{equation}
and
\begin{equation}
n_{s3}=n_2(r_2^-),\qquad
n_{s4}=n_3(r_2^+).
\end{equation}

To investigate the response to the kick-induced perturbation, we introduce a small deformation of the stationary state. In the local-interface approximation, the perturbed wavefunction is written as
\begin{equation}
\widetilde{\Psi}_j
=
\Psi_j
\exp\left[
i\Gamma_j^\pm
e^{\pm k_r(r-r_i)}
\cos(m\theta-\omega t)
\right],
\label{eqn::Excited wavefunction}
\end{equation}
where $\phi_j$ is modified as $\Gamma_j^\pm e^{\pm k_r(r-r_i)}\cos(m\theta-\omega t)$, \(m\) is the integer azimuthal mode number and \(k_r\) is the radial wave number. The sign depends on whether the condensate is inside ($+$) or outside ($-$) of the interface. The exponential radial dependence represents the local planar-interface approximation to the perturbation.

The corresponding interface deformations are taken as
\begin{equation}
R_1=r_1+\eta_1
=r_1+c_1\sin(m\theta-\omega t),
\label{eqn::Curve1}
\end{equation}
and
\begin{equation}
R_2=r_2+\eta_2
=r_2+c_2\sin(m\theta-\omega t),
\label{eqn::Curve2}
\end{equation}
where \(c_1\) and \(c_2\) are small perturbation amplitudes.

The radial velocity of the \(j\)-th superfluid component is determined by the phase gradient,
\begin{equation}
v_{r,j}
=
\frac{\hbar}{m_j}
\frac{\partial\phi_j}{\partial r}.
\label{eqn::radial_velocity}
\end{equation}
The kinematic boundary condition requires the radial velocity of the condensate at the interface to coincide with the interface velocity. For the first interface,
\begin{equation}
\frac{\partial R_1}{\partial t}
=
\left.
\frac{\hbar}{m_j}
\frac{\partial\phi_j}{\partial r}
\right|_{r=R_1},
\qquad j=1,2.
\end{equation}
Using Eq.~\eqref{eqn::Curve1} and retaining only terms linear in the perturbation amplitude gives
\begin{equation}
\pm\frac{\hbar}{m_j}\Gamma_j^\pm k_r
=
-\omega c_1,
\qquad j=1,2.
\label{eqn::Angular velocity1}
\end{equation}
Similarly, for the second interface,
\begin{equation}
\frac{\partial R_2}{\partial t}
=
\left.
\frac{\hbar}{m_j}
\frac{\partial\phi_j}{\partial r}
\right|_{r=R_2},
\qquad j=2,3,
\end{equation}
which gives
\begin{equation}
\pm\frac{\hbar}{m_j}\Gamma_j^\pm k_r
=
-\omega c_2,
\qquad j=2,3.
\label{eqn::Angular velocity2}
\end{equation}

Substituting Eqs.~\eqref{eqn::Excited wavefunction}, \eqref{eqn::Curve1}, and \eqref{eqn::Angular velocity1} into Eq.~\eqref{eqn::Pressure1}, and retaining only terms linear in the perturbation amplitudes, gives
\begin{equation}
\begin{aligned}
&
\left[
f_2-r_1\phi_{d_2}
\cos(2k_{L_2}r_1\cos\theta)\delta t
\right]n_{s2}
+\frac{\rho_2}{k_r}\omega^2
\\
={}&
\left[
f_1-r_1\phi_{d_1}
\cos(2k_{L_1}r_1\cos\theta)\delta t
\right]n_{s1}
-\frac{\rho_1}{k_r}\omega^2
+\frac{\sigma_1m^2}{r_1^2},
\end{aligned}
\label{eqn::Angular velocity3}
\end{equation}
where
\begin{equation}
\rho_1=m_1n_{s1},
\qquad
\rho_2=m_2n_{s2},
\end{equation}
and
\begin{equation}
f_1=U_1'(r_1+\eta_1),
\qquad
f_2=U_2'(r_1+\eta_1).
\end{equation}

Similarly, substituting Eqs.~\eqref{eqn::Excited wavefunction}, \eqref{eqn::Curve2}, and \eqref{eqn::Angular velocity2} into Eq.~\eqref{eqn::Pressure2} gives
\begin{equation}
\begin{aligned}
&
\left[
f_4-r_2\phi_{d_3}
\cos(2k_{L_3}r_2\cos\theta)\delta t
\right]n_{s4}
+\frac{\rho_3}{k_r}\omega^2
\\
={}&
\left[
f_3-r_2\phi_{d_2}
\cos(2k_{L_2}r_2\cos\theta)\delta t
\right]n_{s3}
-\frac{\rho_2'}{k_r}\omega^2
+\frac{\sigma_2m^2}{r_2^2},
\end{aligned}
\label{eqn::Angular velocity4}
\end{equation}
where
\begin{equation}
\rho_2'=m_2n_{s3},
\qquad
\rho_3=m_3n_{s4},
\end{equation}
and
\begin{equation}
f_3=U_2'(r_2+\eta_2),
\qquad
f_4=U_3'(r_2+\eta_2).
\end{equation}

Introducing the effective force-density differences
\begin{equation}
\begin{aligned}
F_1={}&
n_{s2}
\left[
f_2-r_1\phi_{d_2}
\cos(2k_{L_2}r_1\cos\theta)\delta t
\right]
\\
&-
n_{s1}
\left[
f_1-r_1\phi_{d_1}
\cos(2k_{L_1}r_1\cos\theta)\delta t
\right],
\end{aligned}
\label{eqn::F1}
\end{equation}
and
\begin{equation}
\begin{aligned}
F_2={}&
n_{s4}
\left[
f_4-r_2\phi_{d_3}
\cos(2k_{L_3}r_2\cos\theta)\delta t
\right]
\\
&-
n_{s3}
\left[
f_3-r_2\phi_{d_2}
\cos(2k_{L_2}r_2\cos\theta)\delta t
\right],
\end{aligned}
\label{eqn::F2}
\end{equation}
the corresponding linear dispersion relations for the two interfaces are
\begin{equation}
\omega_1^2
=
\frac{
\sigma_1 k_r m^2/r_1^2-F_1k_r
}{
\rho_1+\rho_2
},
\label{eqn::Frequency1}
\end{equation}
and
\begin{equation}
\omega_2^2
=
\frac{
\sigma_2 k_r m^2/r_2^2-F_2k_r
}{
\rho_2'+\rho_3
}.
\label{eqn::Frequency2}
\end{equation}

The first terms in the numerators of Eqs.~\eqref{eqn::Frequency1} and \eqref{eqn::Frequency2} represent the restoring effects of the interfacial tensions, while the terms involving $F_1$ and $F_2$ describe the corresponding effective force-density differences across the two interfaces. An interface becomes linearly unstable when its corresponding squared frequency becomes negative, i.e., $\omega_i^2<0$. Thus, the onset of instability is determined by the condition $\omega_i^2=0$, yielding the critical force-density differences
\begin{equation}
F_{1,\mathrm{c}}=\frac{\sigma_1m^2}{r_1^2},
\qquad
F_{2,\mathrm{c}}=\frac{\sigma_2m^2}{r_2^2}.
\end{equation}
Therefore, the first and second interfaces become linearly unstable when
\begin{equation}
F_1>\frac{\sigma_1m^2}{r_1^2},
\qquad
F_2>\frac{\sigma_2m^2}{r_2^2},
\end{equation}
respectively. The equality in each case corresponds to the neutral stability boundary.

For a circular interface, the integer \(m\) determines the azimuthal mode structure. The corresponding physical tangential wave number at the \(i\)-th interface is
\begin{equation}
k_{\parallel,i}=\frac{m}{r_i}.
\label{eqn::tangential_wave_number}
\end{equation}
Thus, the capillary contribution to the dispersion relation can be written as
\begin{equation}
\omega_{c,i}^{\,2}
=
\frac{\sigma_i k_r k_{\parallel,i}^{\,2}}
{\rho_{i}+\rho_{i+1}}.
\label{eqn::capillary_general}
\end{equation}
In the local-interface limit, where the radial and tangential wave numbers are comparable, \(k_r\simeq k_{\parallel,i}\), this reduces to
\begin{equation}
\omega_{c,i}^{\,2}
\simeq
\frac{\sigma_i}
{\rho_{i}+\rho_{i+1}}
k_{\parallel,i}^{3},
\end{equation}
and therefore
\begin{equation}
\omega_{c,i}\propto k_{\parallel,i}^{3/2}.
\label{eqn::capillary_scaling}
\end{equation}
Hence, the apparent dependence on both \(k_r\) and \(m\) in the circular geometry is a consequence of resolving the perturbation into radial and azimuthal components; in the local planar-interface limit, the conventional capillary-wave scaling is recovered.

The short-duration kick provides an impulsive perturbation to the interfaces and produces a finite change in their velocities. Accordingly, the RMI is characterized by the velocity jump
\begin{equation}
\Delta\dot R_i
=
\dot R_i(0^+)-\dot R_i(0^-),
\end{equation}
which acts as the initial condition for the subsequent evolution of the interfacial perturbation. The dispersion relations above describe the linear response of the interfaces, while the subsequent nonlinear development of the perturbations is obtained from the time-dependent GP dynamics. This variational treatment is consistent with the established description of externally driven BEC interfaces and pulse-induced RMI~\cite{PhysRevA.83.043623}.

\section{Numerical Results}
\label{Numerical}

\begin{figure}[t]
    \centering
    \includegraphics[width=0.9\linewidth]{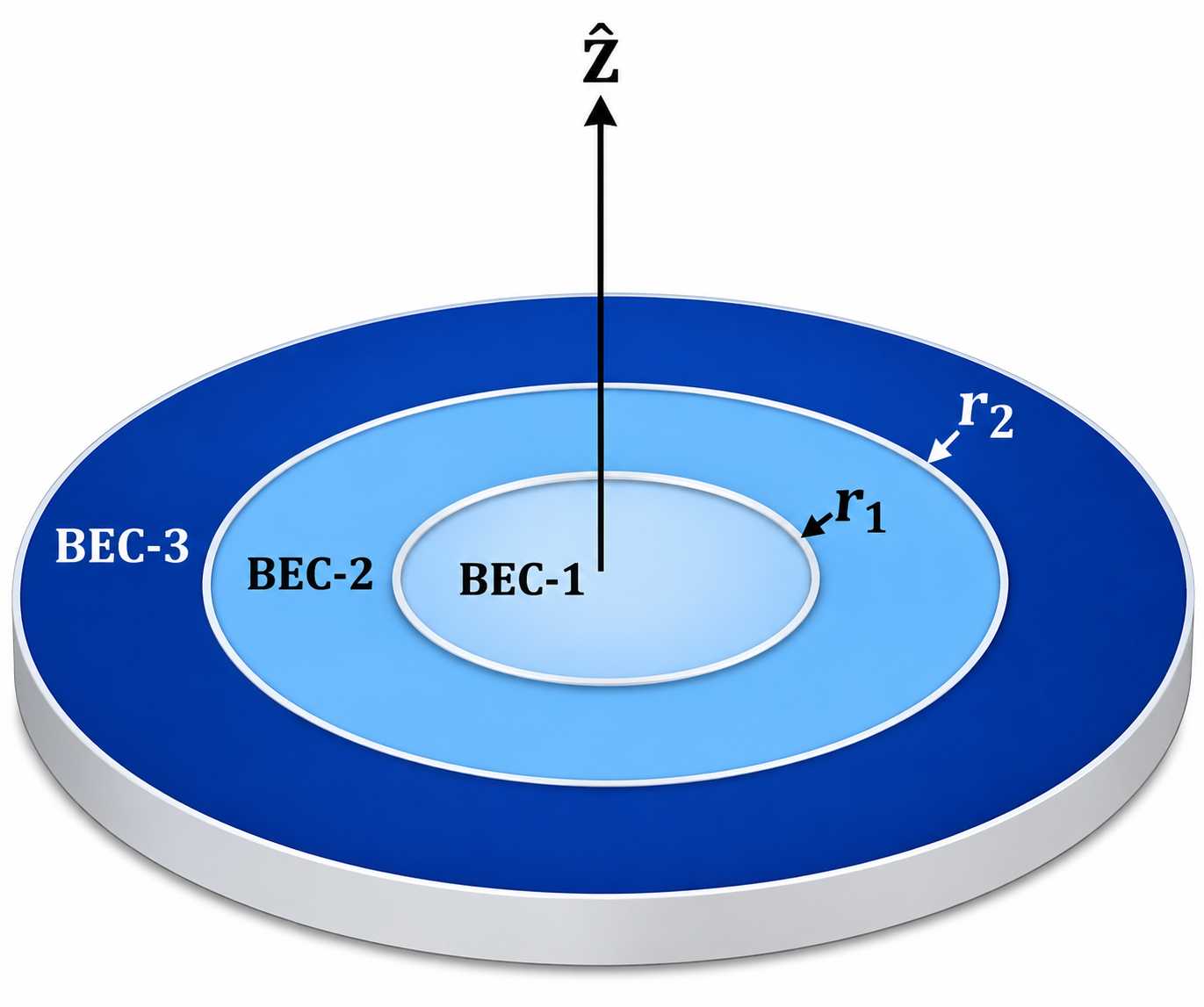}
    \caption{ Schematic representation of a phase-separated three-component Bose-Einstein condensate in a quasi-2D harmonic trap.}
    \label{fig:1}
\end{figure}

We study a three-component BEC made of a \(^{87}\text{Rb}\)-\(^{87}\text{Rb}\)-\(^{87}\text{Rb}\) mixture, trapped in a quasi-2D harmonic potential. Using split-step Crank-Nicolson~\cite{crank_practical_1947,antoine_computational_2013,muruganandam_fortran_2009} method in imaginary time propagation, we solved Eq.~\ref{eqn::RNLS} numerically to find the ground state of a phase-separated three-component BEC. To see the dynamical evolution of the system, we use the ground state as initial state and solve the Eq.~\ref{eqn::RNLS} in real time propagation. In our simulations, the spatial domain spans \( -20l \) to \( 20l \) along \( x \) and \( y \), using an \( 801 \times 801 \) grid. Here \( l = \sqrt{\hbar/m_j \omega_r} \) is the oscillator length. The resolution is \( \Delta x = \Delta y = 0.05l \), with a time step of \( \delta t = (2 \times 10^{-4})/\omega \).
The trap has a frequency of \(\omega_r = 2\pi \times 50\,\text{Hz}\) and an aspect ratio of \(\lambda = 50\). In this setup~\cite{PhysRevA.108.013320,PhysRevLett.101.040402,PhysRevLett.88.093201,PhysRevA.80.050701,1kgv-h6pw,bougas_observation_2026}, where BEC-2 $(|F = 2, m_F = 0\rangle)$ is positioned between BEC-1 $(|F = 1, m_F = -1\rangle)$ and BEC-3 $(|F = 1, m_F = 0\rangle)$, as shown in Fig.~\ref{fig:1}.

We take the intraspecies \(s\)-wave scattering lengths as \(a_{11} = 92.4a_0\), \(a_{22} = 94.5a_0\) and \(a_{33} = 100.4a_0\), where \(a_0\) is the Bohr radius. To keep the condensates in the immiscible regime, the interspecies scattering lengths are set as \(a_{12} = a_{21} = 213a_0\), \(a_{13} = a_{31} = 213a_0\), and \(a_{23} = a_{32} = 127a_0\)~\cite{PhysRevA.108.013320}. Using these parameter, the ground state is found by numerically solving the Gross-Pitaevskii equation, Eq.~\ref{eqn::RNLS}. We take large interspecies interaction $a_{13}=213a_0$ to make BEC-1 and BEC-3 strongly immiscible, while the asymmetric interspecies scattering length of BEC-2 ($a_{12}=213a_0$, $a_{23}=127a_0$) prefer its position between them. Together with $a_{11}<a_{22}<a_{33}$, this leads to a stable nested shell structure with BEC-1 in the core, BEC-2 in the intermediate shell, and BEC-3 in the outer region.

\begin{figure}[t]
    \centering
    \includegraphics[width=1\linewidth]{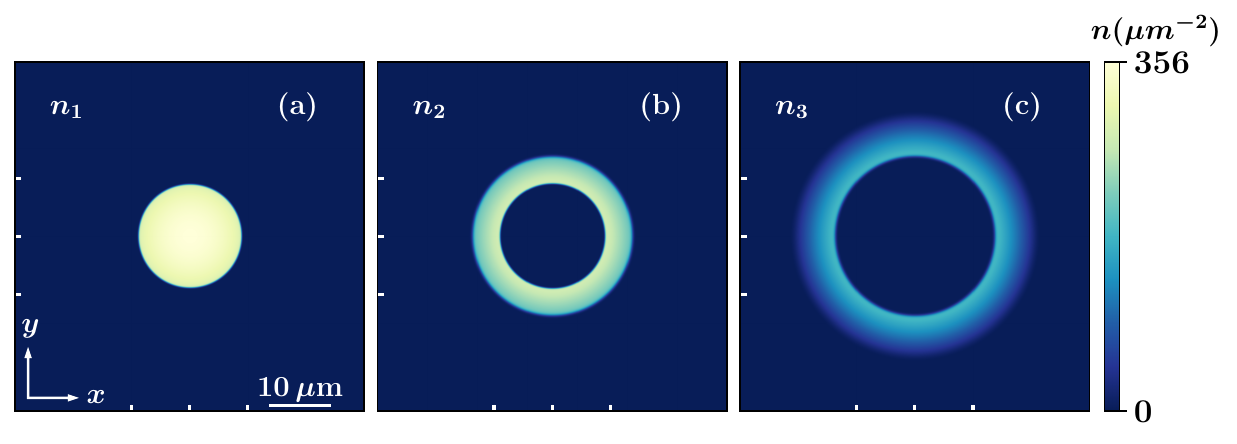}
    \caption{ Ground state density profile of the three components in a phase-separated Bose-Einstein condensate. The system is confined in an anisotropic trap with $(\omega_r, \omega_z) = 2\pi \times (50,2500)$ Hz, and each component contains $N_j = 80,000$ atoms. The intracomponent scattering lengths are $a_{11}=92.4a_0$, $a_{22}=94.5a_0$, and $a_{33}=100.4a_0$, while the intercomponent scattering lengths are $a_{12}=a_{21}=213a_0$, $a_{13}=a_{31}=213a_0$, and $a_{23}=a_{32}=127a_0$, where $a_0$ denotes the Bohr radius.}
    \label{fig:2}
\end{figure}

\begin{figure}[b]
    \centering
    \includegraphics[width=1\linewidth]{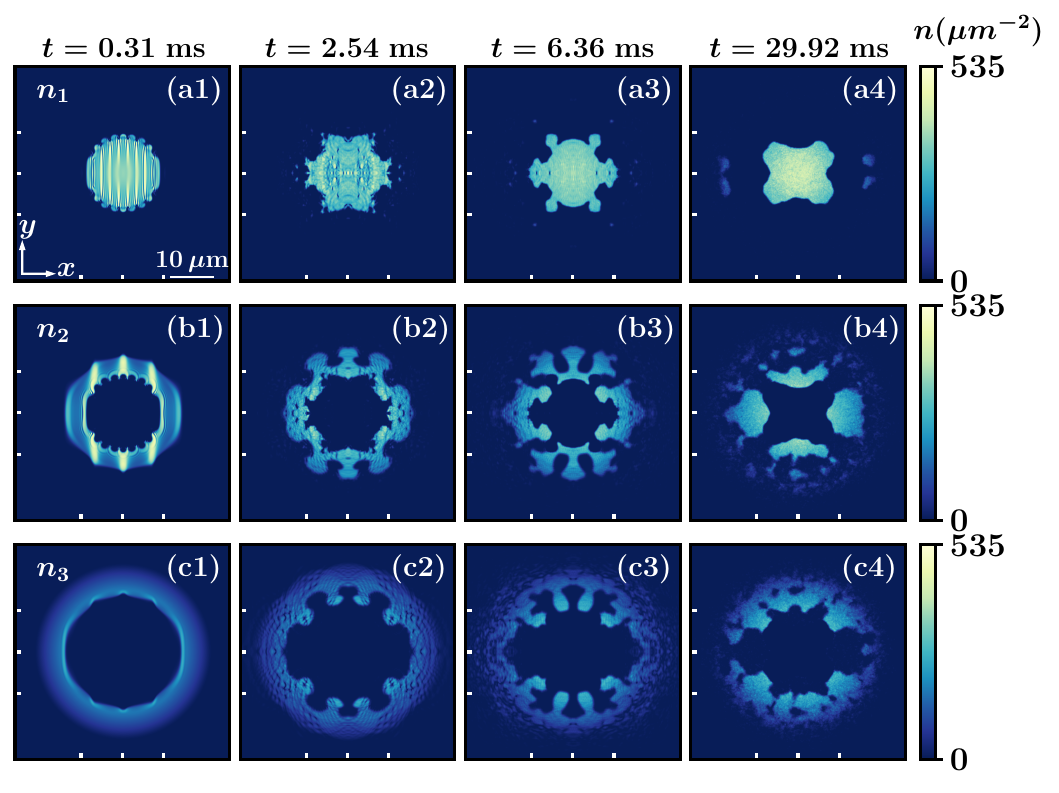}
    \caption{ Time dynamics of a three-component Bose-Einstein condensate confined in a quasi-2D harmonic trap, following driving parameters applied to BEC-1 ($\phi_{d_1}=8$, $k_{L_1}=2.3$), BEC-2 ($\phi_{d_2}=6$, $k_{L_2}=0.7$), and BEC-3 ($\phi_{d_3}=0$, $k_{L_3}=0$). Panels (a1)-(a4), (b1)-(b4), and (c1)-(c4) show the time evolution of BEC-1, BEC-2, and BEC-3, respectively. Density snapshots at $t = 0.31$ ms, $2.54$ ms, $6.36$ ms, and $29.92$ ms illustrate the interfacial deformation – including spikes, bubbles, and mushroom patterns – that characterises the development of the RMI.}
    \label{fig:3}
\end{figure}

We begin our analysis from the stationary state shown in Fig.~\ref{fig:2}. An impulsive kick induced potential is then applied to excite the system and induce the RMI in the three-component Bose-Einstein condensate. Specifically, the driving parameters are chosen as $\phi_{d_1}=8$ and $k_{L_1}=2.3$ for BEC-1, $\phi_{d_2}=6$ and $k_{L_2}=0.7$ for BEC-2, while no driving is applied to BEC-3, i.e., $\phi_{d_3}=0$ and $k_{L_3}=0$. For these parameters, the instantaneous trapping potential can locally acquire an anti-trapping character, a regime that has also been explored in previous studies~\cite{PhysRevE.106.014201, PhysRevA.66.063602, PhysRevA.70.053617, PhysRevLett.94.050402, doi:10.1126/science.1071021}. In the present work, however, the driving is applied through Gaussian kicks of the form $V(t)\propto\exp(-t^{2}/\tau^{2})$, with a pulse width of $\tau=0.09$ ms. Consequently, the anti-trapping contribution exists only for a small duration and does not produce any appreciable expansion or destabilization of the condensate. Instead, it acts as an impulsive perturbation that transfers momentum while the overall harmonic confinement remains effective throughout the evolution.

Using the numerically obtained equilibrium densities and interface radii described in Sec.~\ref{Theory-C}, we estimate the interfacial tensions to be $\sigma_{1} = 6.73 \times 10^{-16} $ N/m and $\sigma_{2} = 1.82 \times 10^{-16} $ N/m following Refs.~\cite{PhysRevA.82.063604,PhysRevA.66.013612,PhysRevA.78.023624,PhysRevA.102.033320}. The corresponding capillary-wave frequencies are $\omega_{c1}=282.69~\mathrm{s^{-1}}$ and $\omega_{c2}=46.92~\mathrm{s^{-1}}$ for the inner and outer interfaces, respectively. Although the driving pulse is not an ideal delta function, its duration is much shorter than the characteristic capillary-wave timescale, satisfying the condition $\omega_c\tau\ll1$ for both interfaces~\cite{PhysRevA.82.043608}. Therefore, the applied excitation is effectively impulsive, and the resulting interfacial dynamics can be unambiguously identified as the RMI.

The system's real-time progression is depicted in Fig.~\ref{fig:3}. At $t = 0.31~\mathrm{ms}$ [Figs.~\ref{fig:3}(a1),~\ref{fig:3}(b1), ~\ref{fig:3}(c1)], Faraday like patterns~\cite{rj7v-q9gy} are visible in the density profile of the condensate and the secondary KHI~\cite{PhysRevA.82.043608} manifests at both boundaries, with spike-like extensions starting to appear at the initial interface, signifying the initiation of axisymmetry disruption caused by the applied kicked potential. At $t = 2.54~\mathrm{ms}$ [Figs.~\ref{fig:3}(a2),~\ref{fig:3}(b2), ~\ref{fig:3}(c2)], these interfacial disturbances amplify considerably, resulting in distinctive mushroom-shaped formations in the second and third components, while bubble-like structures emerge within the condensate. At $t = 6.36~\mathrm{ms}$ [Figs.~\ref{fig:3}(a3),~\ref{fig:3}(b3), ~\ref{fig:3}(c3)], analogous mushroom-like formations emerge in the first component, indicating the full maturation of the RMI. At a subsequent time, $t = 29.92~\mathrm{ms}$ [Figs.~\ref{fig:3}(a4),~\ref{fig:3}(b4), and~\ref{fig:3}(c4)], significant distortions of both interfaces, accompanied by the formation of quantised vortices, are observed, indicating the emergence of quantum turbulence (QT).

\section{Quantum Turbulence}
\label{QT}
We next turn our attention to the kinematics of the flow by examining the kinetic-energy spectrum obtained from the Helmholtz decomposition of the density-weighted velocity field into incompressible and compressible components. The incompressible component is associated primarily with quantised vortices, whereas the compressible component captures sound-wave excitations~\cite{PhysRevLett.78.3896, doi:10.1073/pnas.1312737110}. The corresponding spectra in Fourier space, obtained using Parseval's theorem, provide information about the dominant flow structures, characteristic scaling regimes, and redistribution of kinetic energy across spatial scales~\cite{PhysRevLett.94.065302, PhysRevA.86.053621}.

We introduce the density weighted velocity $\sqrt{n_j}\mathbf{u}_j$ for each condensate component. Here $\mathbf{u}_j=(\hbar/m_j)\nabla\phi_j$ is the superfluid velocity, $n_j$ and $\phi_j$ are the local density and phase of the $j$-th condensate respectively. The Helmholtz decomposition then expresses the velocity field as the sum of a divergence free part $\mathbf{u}_j^{\mathrm{ic}}$ and a curl free part $\mathbf{u}_j^{\mathrm{c}}$ such that $\mathbf{u}_j=\mathbf{u}_j^{\mathrm{ic}}+\mathbf{u}_j^{\mathrm{c}}$, $\nabla\cdot\mathbf{u}_j^{\mathrm{ic}}=0$, $\nabla\times\mathbf{u}_j^{\mathrm{c}}=0$.

The corresponding kinetic energies are then calculated according to
\begin{equation}
\mathcal{E}_j^{\mathrm{ic(c)}}=
\frac{1}{2}\int n_j
\left|\mathbf{u}_j^{\mathrm{ic(c)}}\right|^2\,dx\,dy.
\end{equation}

The spectral distribution of these two contributions is obtained in Fourier space as~\cite{PhysRevA.106.043322, PhysRevA.81.063630, PhysRevA.80.023618}
\begin{equation}
E_j^{\mathrm{ic(c)}}(k)
=
\frac{k}{2}
\sum_{q=x,y}
\int_{0}^{2\pi}
\left|
F_q\left(\sqrt{n_j}\,u_{q,j}^{\mathrm{ic(c)}}\right)
\right|^2\,d\phi,
\end{equation}
where $F_q$ denotes the Fourier transform of the $q$-th Cartesian component of the density-weighted velocity field $\sqrt{n_j}\mathbf{u}_j$, with $\mathbf{u}_j=(u_{x,j},u_{y,j})$.

As the system evolves toward the turbulent state, fluctuations span a wide range of characteristic length scales. From the spectral analysis we identify two characteristic length scales, the Thomas-Fermi radius $R_{\mathrm{TF_j}}=\sqrt{2\mu_j},$ and the condensate healing length $\xi_j=\frac{1}{\sqrt{\mu_j}},$ where $\mu_j$ is the chemical potential of the $j$-th component~\cite{10.1063/5.0190917}. We also introduce the corresponding characteristic wave numbers $k_{\xi_j}=\frac{2\pi}{\xi_j}$, $
k_{\mathrm{TF_j}}=\frac{2\pi}{R_{\mathrm{TF_j}}}
$.

\begin{figure}[t]
    \centering
    \includegraphics[width=1\linewidth]{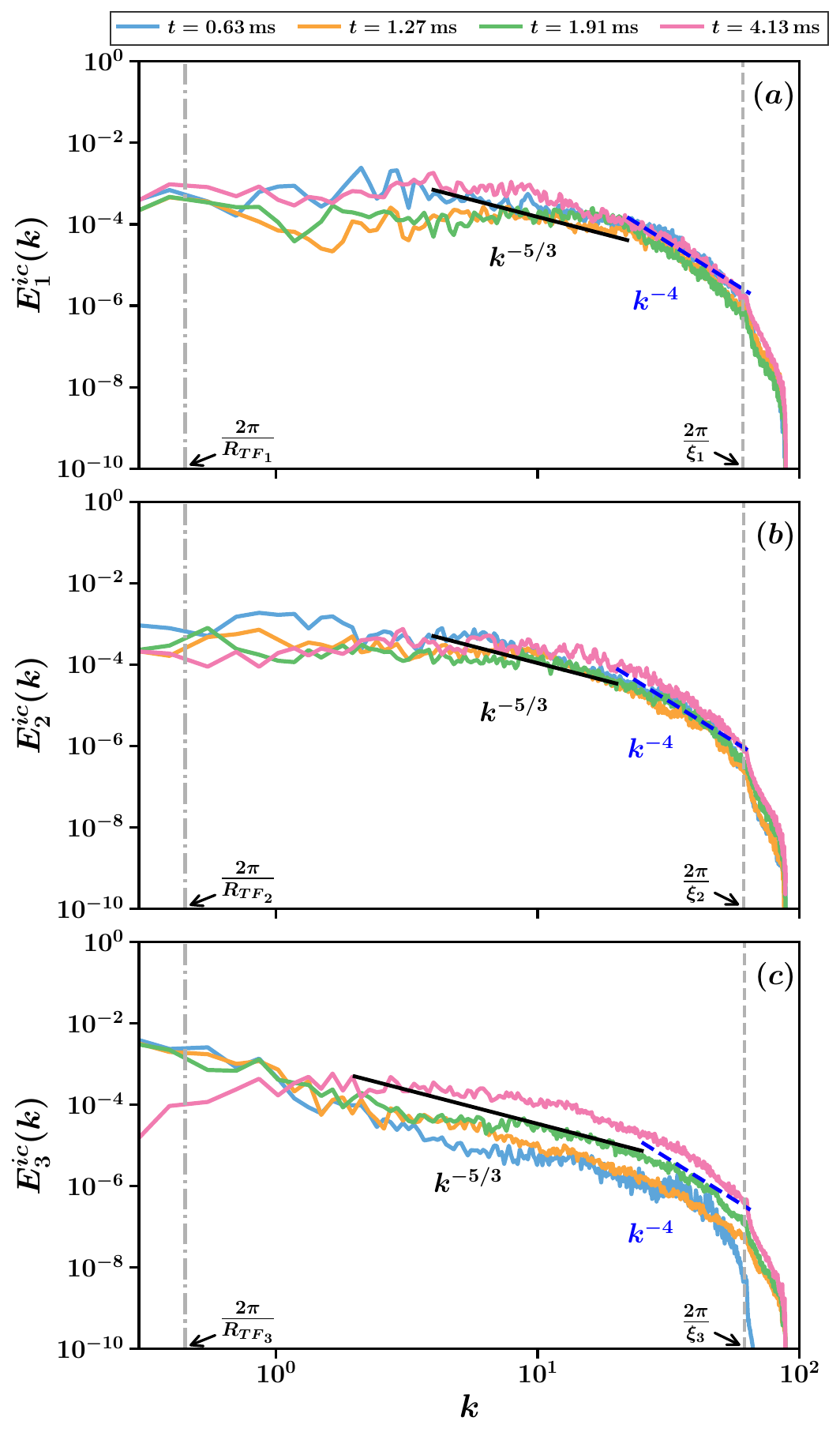}
    \caption{ The incompressible kinetic energy spectra of (a) BEC-1, (b) BEC-2 and (c) BEC-3 at different instants of time for RMI. The vertical dashed lines indicate the characteristic wave numbers, $k_{\mathrm{TF_j}}=\frac{2\pi}{R_{\mathrm{TF_j}}}$ and $k_{\xi_j}=\frac{2\pi}{\xi_j}$, corresponding to the Thomas-Fermi radius $R_{\mathrm{TF_j}}=\sqrt{2\mu_j}$, and the healing length, $\xi_j=\frac{1}{\sqrt{\mu_j}}$  respectively, of the j-th component.}
    \label{fig:4}
\end{figure}

\begin{figure}[t]
    \centering
    \includegraphics[width=1\linewidth]{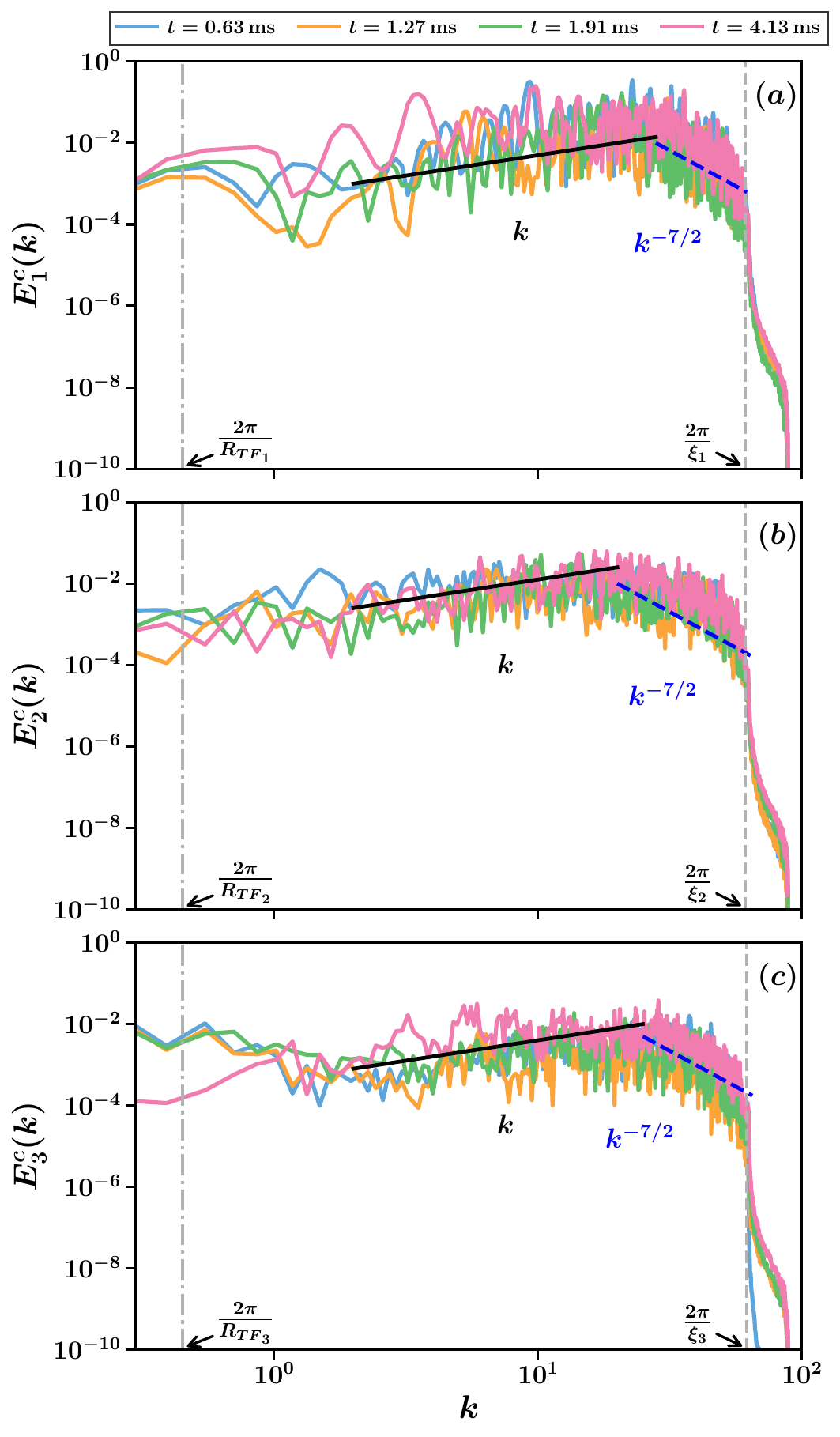}
    \caption{ The compressible kinetic energy spectra of (a) BEC-1, (b) BEC-2 and (c) BEC-3 at different instants of time for RMI. The vertical dashed lines indicate the characteristic wave numbers, $k_{\mathrm{TF_j}}=\frac{2\pi}{R_{\mathrm{TF_j}}}$ and $k_{\xi_j}=\frac{2\pi}{\xi_j}$, corresponding to the Thomas-Fermi radius $R_{\mathrm{TF_j}}=\sqrt{2\mu_j}$, and the healing length, $\xi_j=\frac{1}{\sqrt{\mu_j}}$  respectively, of the j-th component.}
    \label{fig:5}
\end{figure}

We first examine the incompressible kinetic-energy spectra. During the evolution, the incompressible kinetic-energy spectrum of each condensate component, $E_{\mathrm{kin},j}^{\mathrm{ic}}(k)$, exhibits two distinct scaling regimes characterized by power-law behavior~\cite{GHOSHDASTIDAR2022127776, PhysRevA.80.023618, PhysRevA.106.023306, pj6n-3w6p}. For the first component, the spectrum follows the Kolmogorov scaling $E_{\mathrm{kin},1}^{\mathrm{ic}}(k)\propto k^{-5/3}$ over the range $(4<k<22)$, indicating the turbulent transfer of incompressible kinetic energy across large length scales [see Fig.~\ref{fig:4}(a)]. At higher wave numbers $(22<k<65)$, the Kolmogorov regime is followed by the Vinen turbulence regime, where the incompressible kinetic-energy spectrum exhibits a $k^{-4}$ scaling [see Fig.~\ref{fig:4}(a)]. Similarly, the second component exhibits a $k^{-5/3}$ scaling over $4<k<20$, followed by a $k^{-4}$ scaling in the range $20<k<65$ [see Fig.~\ref{fig:4}(b)]. For the third component, the low-wave-number region $(2<k<25)$ is well described by the $k^{-5/3}$ scaling, whereas the high-wave-number region $(25<k<65)$ follows a $k^{-4}$ power law [see Fig.~\ref{fig:4}(c)].

The kinetic energy spectra further demonstrates a reassignment of incompressible kinetic energy. As the system approaches turbulence, the spectral energy increases at low wave numbers $(0.4<k<1)$ while decreasing around intermediate wave numbers $(k\approx20)$. This transfer of spectral weight from higher to lower wave numbers is a signature of an inverse energy cascade, demonstrating the transport of incompressible kinetic energy from smaller to larger length scales.

Next, we consider the behavior of the compressible component of the kinetic-energy spectrum, shown in Fig.~\ref{fig:5}. The time-averaged spectrum exhibits distinct scaling regimes over different wave-number intervals. At low wave numbers, the spectrum approximately follows a linear scaling, $E_{\mathrm{kin},j}^{\mathrm{c}}(k)\propto k$, indicating that long-wavelength compressible excitations contribute significantly to the large-scale dynamics of all three components. At larger wave numbers, the spectrum crosses over to a steeper power-law behavior, approximately $E_{\mathrm{kin},j}^{\mathrm{c}}(k)\propto k^{-7/2}$. The crossover occurs near the wave number associated with the healing length, suggesting a change in the character of the compressible dynamics at smaller length scales. In decaying quantum turbulence, compressible kinetic energy can be redistributed toward higher wave numbers as the system evolves toward a more thermalized state. In the present conservative GP dynamics, however, no explicit dissipative mechanism or coupling to a thermal reservoir is included, and therefore complete thermal equilibrium is not expected. Nevertheless, kinetic energy can be exchanged continuously between the incompressible and compressible sectors through vortex-acoustic interactions. Since the condensate is initially vortex-free, quantized vortices emerge dynamically through the nonlinear deformation and collision of the condensate domains, accompanied by the generation of acoustic excitations. The persistent contribution of the compressible kinetic energy throughout the evolution indicates that the acoustic field acts as an important energy reservoir coupled to the vortex dynamics. The observed $k^{-7/2}$ scaling at large wave numbers may therefore be associated with the coupled vortex-acoustic dynamics and the resulting redistribution of compressible kinetic energy across spatial scales.

\section{Conclusions}
\label{Conclusions}

In this work, we have investigated the RMI in a three-component BEC by applying a Gaussian impulsive perturbation that acts as a shock wave. Starting from the coupled GP equations, we developed a hydrodynamic description to analyze the onset and nonlinear evolution of the instability. We demonstrated that the impulsive excitation of the interfaces between condensate components generates the characteristic growth of spikes, bubbles, and mushroom-like structures, in close analogy with the classical hydrodynamic RMI.

Our results further demonstrate that three-component condensates provide a significantly richer platform for studying interfacial hydrodynamics than binary mixtures. The presence of two independently controllable interfaces enables coupled interface dynamics, enhanced control over instability development, and a broader range of emergent nonlinear patterns. This extra degree of freedom allows different instabilities to interact within the same quantum fluid, which cannot be achieved in standard two-component condensates. 

Finally, the simultaneous observation of the Kolmogorov ($k^{-5/3}$), Saffman ($k^{-4}$), and compressible ($k$ and $k^{-7/2}$) scaling regimes in the kinetic-energy spectra provides strong evidence for the coexistence of classical-like vortex cascades and intrinsically quantum acoustic excitations in turbulent Bose-Einstein condensates.

Our findings demonstrate that ultra-cold atomic gasses provide a tunable quantum platform to explore interfacial instabilities traditionally studied in various platforms. The expected interface deformation, vortex production, and excitation signatures are all accessible with existing ultracold atom technology, making Rb mixtures especially suitable for observing these effects. The results of our study laid the framework for experimental exploration. Future work can explore how these results change at finite temperature, in turbulent regimes, or when beyond-mean-field effects are included. Such studies would help establish the RM instability in BECs as a unifying link between quantum fluid behavior and a wide range of classical hydrodynamic instabilities, including Kelvin-Helmholtz~\cite{PhysRevB.81.094517,PhysRevA.82.063604,PhysRevA.97.053608,PhysRevA.104.023312,1kgv-h6pw}, Rayleigh-Plateau~\cite{PhysRevA.107.063312}, Faraday~\cite{PhysRevA.102.033320,PhysRevLett.127.113001}, fingering~\cite{PhysRevA.97.023625}, capillary~\cite{PhysRevA.83.053606}, Crow~\cite{PhysRevA.84.021603}, modulational~\cite{PhysRevA.85.013616,PhysRevA.97.011604}, snake~\cite{PhysRevA.93.033618,PhysRevA.88.043639,PhysRevA.110.L061303}, and Rosensweig~\cite{PhysRevLett.102.230403} instabilities.

\section{Acknowledgements}
The author expresses gratitude to PARAM Shakti at the Indian Institute of Technology Kharagpur, a National Supercomputing Mission initiative of the Government of India, for providing the computational resources essential for this research. The author also gratefully acknowledges the financial support provided by the UGC Fellowship, India.

\bibliographystyle{apsrev4-2}
\bibliography{ReferencesRM}

\end{document}